\documentclass[twocolumn,prb,amsmath,amsfonts,showpacs,floatfix]{revtex4}
\usepackage{graphicx}
\usepackage{epsf}

\begin{document}
\title{Linear and nonlinear Stark effect in triangular molecule}

\author{Bogdan R. Bu{\l}ka$^1$, Tomasz Kostyrko$^2$ and Jakub {\L}uczak$^1$ }
\affiliation{$^1$Institute of Molecular Physics, Polish Academy of
Science, ul. M. Smoluchowskiego 17, 60-179 Pozna{\'n}, Poland}
\affiliation{$^2$Faculty of Physics, A. Mickiewicz University,
ul.~Umultowska~85,61-614~Pozna\'{n},~Poland}

\date{\today}

\begin{abstract}
We analyze changes of the electronic structure of a triangular molecule under the influence of an electric field (i.e., the Stark effect). The effects of the field are shown to be anisotropic and include both a linear and a nonlinear part. For strong electron correlations, we explicitly derive exchange couplings in an effective spin Hamiltonian. For some conditions one can find a dark spin state, for which one of the spins is decoupled from the others. The model is also applied for studying electronic transport through a system of three coherently coupled quantum dots. Since electron transfer rates are anisotropic, the current characteristics are anisotropic as well, differing for small and large electric field.
\end{abstract}


\maketitle

\section{Introduction}

In this paper we investigate electronic properties of a
model of a triangular molecule in a presence of an electric
field. Recently, similar models were considered for systems
of three coherently coupled quantum dots (QDs), \cite{Gardreau2006,Delgado2008,Busl,Brandes,Emary,Poltl,Kostyrko09,Zitko},
for magnetic interactions in molecules,\cite{gatteschi,choi,trif,tsukerblatt}
as well as for complex phase orderings in strongly correlated electronic
materials with a triangular lattice (e.g., multiferroics \cite{bulaevskii},
cobaltates \cite{cobaltates}, or organic compounds \cite{powell}). These simple
models exhibit plenty of interesting physics. For example, in the system of quantum dots, one finds complex charge and spin arrangements\cite{Gardreau2006}, which can be classified according a set of topological Hund's rules.
For some interference conditions a so-called dark state can
occur, for which one of the QDs is decoupled from the
reservoirs and an electron can be trapped.\cite{Brandes}
Consequently, electronic transport is changed; one can observe a rectification
effect, negative differential resistance, and an enhancement of
the shot noise.\cite{Emary} Moreover, one can expect the Aharonov-Bohm
effect, when a magnetic flux penetrates the triangle. Moreover, one can expect theAharonov-Bohm
effect, when a magnetic flux penetrates the triangle. The effect
leads to a crossover from the singlet to the triplet ground
state, which manifests itself in spin blockade in transport~\cite{Delgado2008} and
interesting spin dynamics under two-electron-spin-resonance
\cite{Busl}.  For the regime of coherent transport one can expect the
very rich phase diagram with many types of the Kondo resonances
\cite{Zitko}.

Our studies concern the Stark effect in the system of strongly correlated electrons
and they are addressed mainly to coherently coupled quantum dots. For a small number of electrons
($n=1$ and $n=2$ in the triangle) the electric field induces a large polarization,
and many aspects were already considered~\cite{Gardreau2006,Delgado2008,Busl,Brandes,Emary,Poltl,Kostyrko09,Zitko}.
In this paper we focus on the situation with $n=3$ electrons,
for which the induced polarization is minor,
because the Coulomb interactions dominate and hinder a possible shift of an electronic charge.
First we will show that the
electric field leads to splitting of energy levels as well as to breaking of the symmetry of the
system and changes the symmetry of wave functions.
We demonstrate that the electric field can induce significant changes in a spin arrangement.
It also results
in changes of coupling between spins and different characteristics of
spin-spin correlation functions with respect to an angle the
  electric field forms with the median of the triangle.
In particular we will show conditions for the appearance of the dark spin state. Second
we will show how the Stark effect manifests itself in electronic transport. We
confine ourselves to the case in which at equilibrium the ground state is
singlet or triplet and for an applied bias voltage excited states with
three electrons (doublets and quadruplets) participate in transport.

\section{Influence of electric field on spin states}

\begin{figure}[ht]
\includegraphics[width=0.38\textwidth,bb=150 370 563 625,clip]{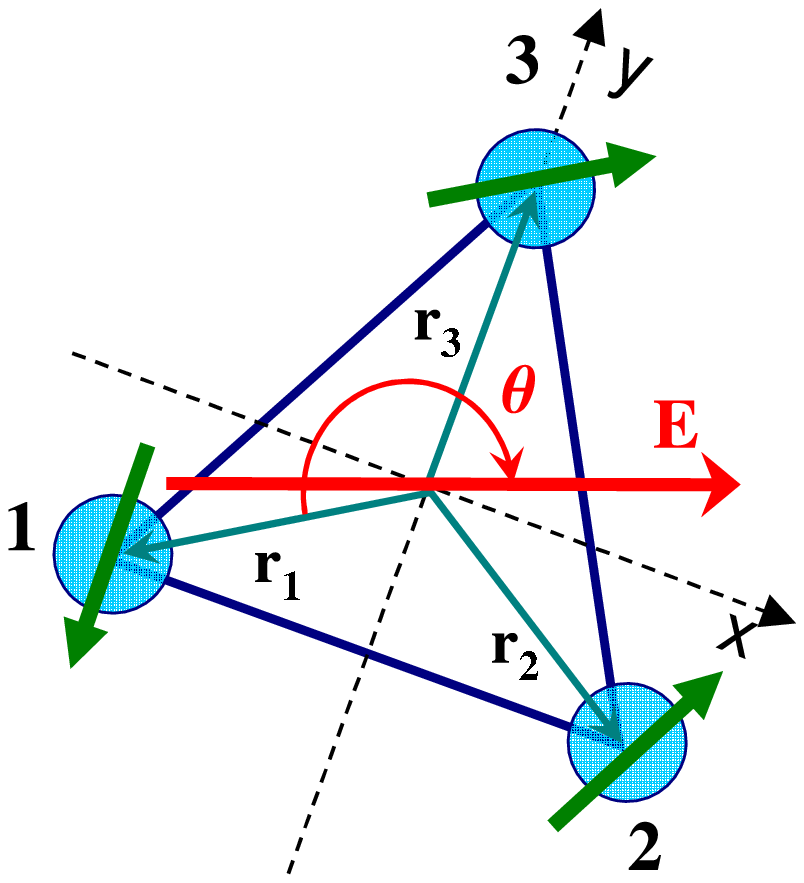}
\caption{(Color online) The considered model of a triangular molecule placed in an
  electric field $\mathbf{E}$.}
\end{figure}

A model of a triangular molecule in the presence of an electric field
$\mathbf{E}$ is shown in Fig.1. The corresponding Hamiltonian can be
expressed as
\begin{eqnarray}\label{model}
  H_M=t\sum_{i<j,\sigma}(c^{\dagger}_{i\sigma}c_{j\sigma}
  +h.c.)+U_0\sum_{i}n_{i\uparrow}n_{i\downarrow}
  \nonumber\\
  +U_1\sum_{i<j,\sigma,\sigma'}n_{i\sigma}n_{j\sigma'}+Eer\sum_{i,\sigma} \cos[(\theta+(i-1)\frac{2\pi}{3}]\,n_{i\sigma},
\end{eqnarray}
where the first term describes electron hopping between nearest neighbor sites.
For the sake of generality we consider $t<0$ as well as $t>0$, for which the model
describes the system of QDs with electrons or holes, respectively.
The second and the third terms correspond to insite and intersite Coulomb
interactions. The last term takes into account the influence of the electric field
$\mathbf{E}$ on the electronic polarization $\hat{\mathbf{P}}=
e\sum_{i,\sigma}\mathbf{r}_i n_{i\sigma}$, where $\mathbf{r}_i$
denotes the vector pointing to the site $i$, $e$ - a charge of an
electron, $\theta$ - the angle between $\mathbf{r}_1$ and
$\mathbf{E}$, $d$ - the length of the arm of the triangle and
$r=d/\sqrt{3}$. One can see that the electric field modulates the local
site energies: $\epsilon_i= g_E \cos[\theta+(i-1)2\pi/3]$. In further
considerations we take $g_E=Eer$ as a coupling parameter.

For $n=3$ electrons in the system the wave functions are constructed
from the singlet and the triplet states by adding an electron (see
[\onlinecite{pauncz}]). There are quadruplet states ($^{3/2}$Q$_{S_z}$) with
$S=3/2$, $S_z=\pm 3/2$, $\pm 1/2$, and the corresponding wave
functions are constructed from the triplet states by adding an
electron. The ground state is, however, the doublet state
($^{1/2}$D$_{S_z}$) with $S=1/2$, $S_z=\pm1/2$, which can be formed
from the states
\begin{eqnarray}\label{f1}
  |\text{D}_{S_z}\rangle_1=\frac{1}{\sqrt{2}}
  c^{\dagger}_{1\sigma}(c^{\dagger}_{2\sigma}c^{\dagger}_{3\overline{\sigma}} -c^{\dagger}_{2\overline{\sigma}}c^{\dagger}_{3\sigma})\left|\mathrm{vac}\right\rangle,\\
\label{f2}
|\text{D}_{S_z}\rangle_2=\frac{1}{\sqrt{6}}[2\;c^{\dagger}_{1\overline{\sigma}} c^{\dagger}_{2\sigma}c^{\dagger}_{3\sigma}
-c^{\dagger}_{1\sigma}(c^{\dagger}_{2\sigma}c^{\dagger}_{3\overline{\sigma}} +c^{\dagger}_{2\overline{\sigma}}c^{\dagger}_{3\sigma})]\left|\mathrm{vac}\right\rangle
\end{eqnarray}
and the states with double site occupancy
$c^{\dagger}_{1\sigma}c^{\dagger}_{1\overline{\sigma}}c^{\dagger}_{2\sigma}\left|\mathrm{vac}\right\rangle$,
$c^{\dagger}_{2\sigma}c^{\dagger}_{2\overline{\sigma}}c^{\dagger}_{3\sigma}\left|\mathrm{vac}\right\rangle$,
etc. The function $|\text{D}_{S_z}\rangle_1$ and
$|\text{D}_{S_z}\rangle_2$ are constructed respectively from the
singlet and the triplet state at the 23--bond by adding an electron to
the QD 1.

\begin{figure}[ht]
\centerline{\epsfxsize=0.45\textwidth \epsfbox{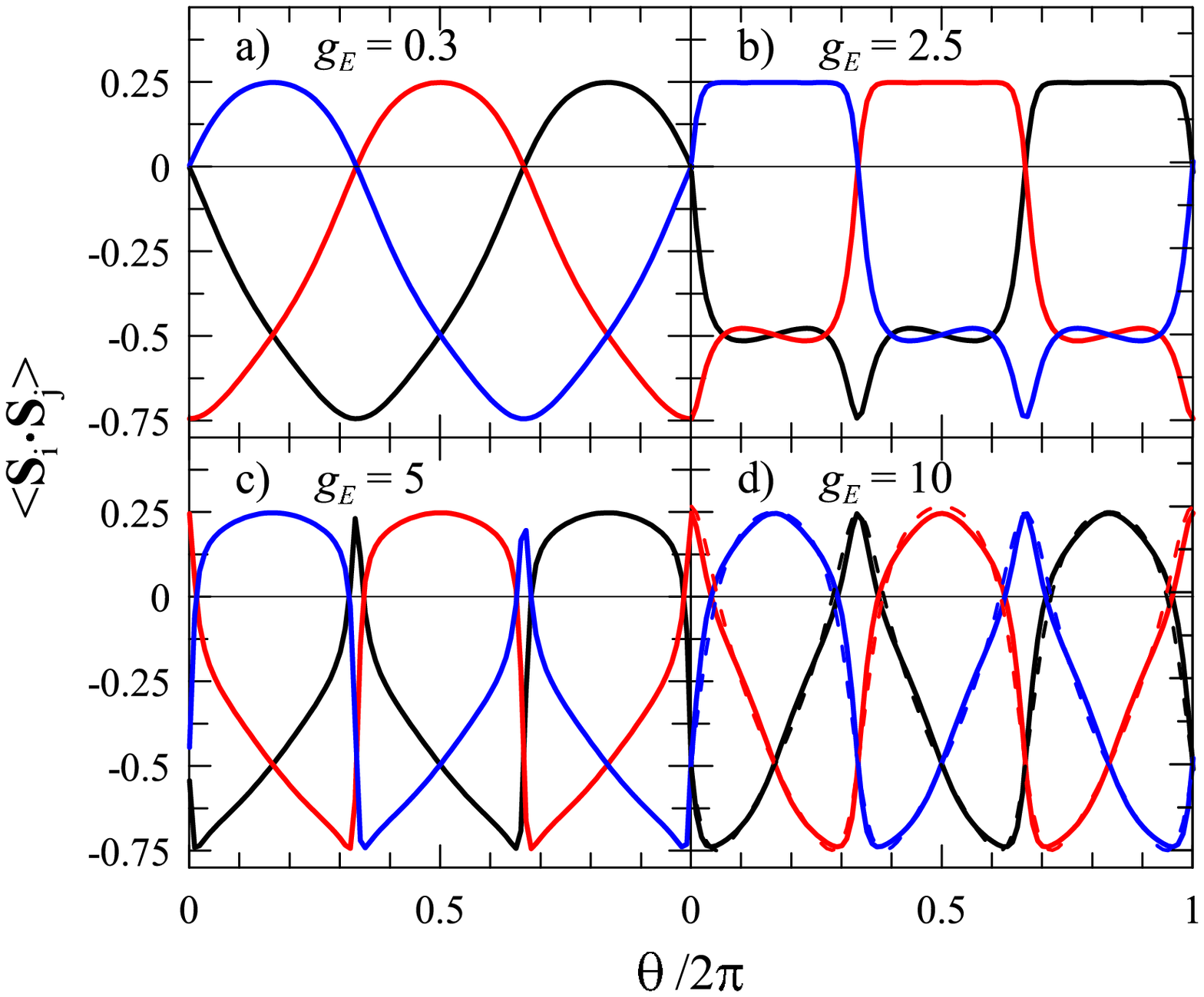} }
\caption{(Color online) Expectation values of the spin correlators
  $\mathbf{S}_1\cdot \mathbf{S}_2$, $\mathbf{S}_2\cdot \mathbf{S}_3$
  and $\mathbf{S}_3\cdot \mathbf{S}_1$ (black, red and blue curves, respectively)
  calculated for the ground state as a function of the angle $\theta$
  of the electric field with respect to the triangle. The plots are
  obtained for the Hubbard model (1) with a large $U_0=30$,
  $U_1=2$, $t=1$, $g_E=0.3$ (a), $g_E=2.5$ (b), $g_E=5$ (c) and
  $g_E=10$ (d). For comparison the dashed curves represent the spin
  correlators for the Heisenberg model (4) with the modulated exchange
  couplings $J_{ij}$ given by Eq.(5) -- [for smaller $g_E\leq 5$ the
  dashed curves cover the solid ones within the plot resolution].}
\end{figure}

Fig.2 presents an evolution of the spin-spin correlation functions
when the electric field increases, it manifests a transition from the
linear to the quadratic Stark effect. For small fields the
spin-correlation functions oscillate with the period $2\pi$. At an
intermediate value $|g_E|=4|t|$ a crossover occurs, and for a larger
$g$ the functions $\langle\mathbf{S}_i\cdot \mathbf{S}_j\rangle$ show
new components due to the quadratic Stark effect. One can see in
Fig.2a and 2b that at $\theta=0$
(where the electric field is perpendicular to the 23-bond and points to the 1-st site) the spin correlators
$\langle\mathbf{S}_1\cdot \mathbf{S}_2\rangle=
\langle\mathbf{S}_3\cdot \mathbf{S}_1\rangle=0$ and
$\langle\mathbf{S}_2\cdot \mathbf{S}_3\rangle\approx -0.75$. This means
that the spin at QD1 is uncoupled from the spins forming the
singlet at the 23-bond, and the corresponding state is
$|\text{D}_{S_z}\rangle_1$. We call such a configuration the {\it
  dark spin state}, in contrast to the dark states for $n=1$ and $n=2$
electrons in the triangle molecule, when their properties are
connected to a specific charge distribution.\cite{Brandes,Emary}
With rotation of the electric field, the dark spin state occurs at
$\theta=2\pi/3$ ($\theta=4\pi/3$), when the uncoupled spin is located
at the QD 2 (3) and the singlet state is at the 12 (31) bond
(the electric field is then perpendicular to the bond). The situation is
more complex for a large $|g_E|>4|t|$, when the quadratic Stark effect begins to
dominate and gradually
changes the period of oscillation of $\langle\mathbf{S}_i\cdot
\mathbf{S}_j\rangle$ as well as the configuration of the dark spin states.

In order to understand the crossover from small to large fields, we perform the
perturbative canonical transformation of the Hubbard
Hamiltonian (1) to an effective Heisenberg Hamiltonian, treating the intersite terms
(both the hopping and the intersite Coulomb interaction terms) as small ones.\cite{macdonald}  To take into
account both the linear and non-linear Stark effect the
perturbation expansion should be carried out to the third
order (see the Appendix for details). For $n=3$ electrons the effective Hamiltonian reads
\begin{eqnarray}\label{ef}
\tilde{H} &=& 3U_1+ \sum_{i<j}J_{ij}
\left(\vec{S}_i\cdot\vec{S}_j- \frac{1}{4}\right)\,,
\end{eqnarray}
with the exchange coupling
\begin{eqnarray}\label{j23e}
J_{ij}= \frac{4t^2}{U_0}+\frac{4t^2(\epsilon_j-\epsilon_i)^2}{U_0^3}
+\frac{8t^3(2\epsilon_m-\epsilon_i-\epsilon_j)}{U_0^3}\,,
\end{eqnarray}
where the indices $i, j,m$ denote three different sites.
Here, we also assumed that the on-site Coulomb interactions are stronger than the electric field, i.e. $U_0\gg{}\epsilon_i$.
It is seen that for the weak field the third order term [the third term in
Eq.(\ref{j23e})] depends linearly on the electric field $E$,
whereas the second order term behaves like a second power of $E$ and
its period of oscillations is twice as large as the linear term.
This result presents one of the main differences between the Stark effect in the system of  strongly correlated electrons and that in atomic physics~\cite{friedrich}.

For a very large $g_E$, when the quadratic term dominates, the dark
spin state can occur for $\mathbf{E}$ parallel to a bond of the
triangle. Then the singlet state is formed on this bond and the
uncoupled spin is at the opposite QD. In this limit there is a direct
exchange process, which is symmetric with respect to exchange of spins
between QDs, and it does not depend on the orientation of the dipole
(it is a quadratic dependence on $E$). Since the exchange
coupling $J_{ij}$ depends on the difference of site energies [see Eq.(\ref{j23e})],
 its maximal value is at the electric field parallel to the bond. For this case two other exchange
couplings are equal, and the dark spin state appears.

\begin{figure}[ht]
\centerline{\epsfxsize=0.35\textwidth \epsfbox{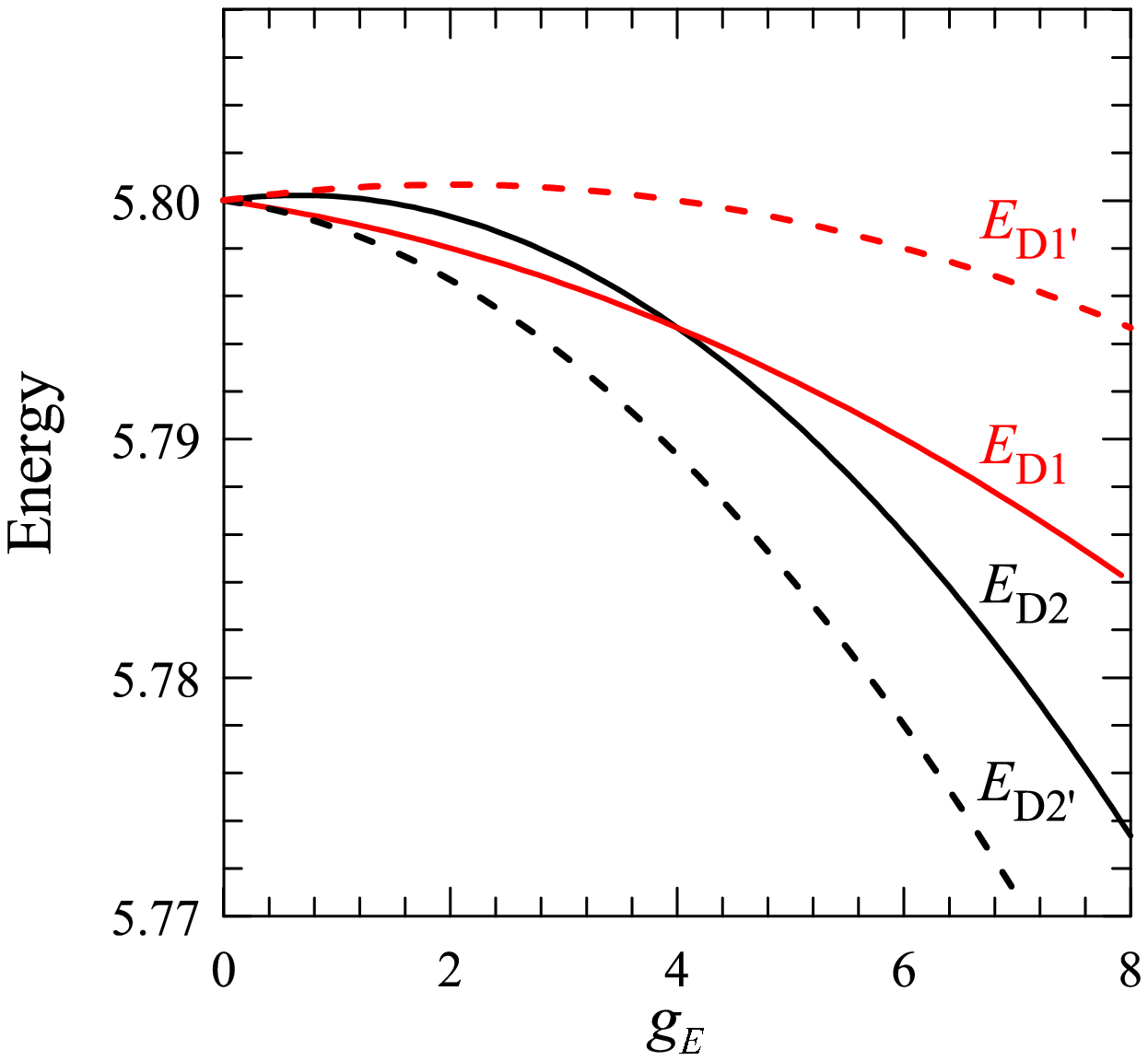} }
\caption{(Color online) Splitting of states by the electric
  field in the effective Heisenberg model (\ref{ef}). The solid curves
  represent the states $E_{\text{D}1}$ and $E_{\text{D}2}$
  for $\theta=0$, whereas the dashed curves represent $E_{\text{D}1'}$,
  $E_{\text{D}2'}$ for $\theta=\pi/3$ ($t=1$, $U_0=30$, $U_1=2$).}
\end{figure}

The eigenvalues of the Hamiltonian (\ref{ef}) can be readily
  obtained as
\begin{eqnarray}
E_{\text{Q}}=3U_1,\\E_{\text{D}1,2}=3U_1-(6J\pm2\Delta)/4
\end{eqnarray}
for quadruplet and doublet, respectively. Here, $J=(J_{12}+J_{23}+J_{31})/3$ and
$\Delta=\sqrt{J_{12}^2+J_{23}^2+J_{31}^2-J_{12}J_{23}-J_{12}J_{31}-J_{23}J_{31}}$.
If the electric field is perpendicular to the $ij$-bond, $\epsilon_i=\epsilon_j$, the exchange couplings $J_{im}=J_{jm}$ and we have the dark spin state.
In particular for $\theta=0$,  $J_{12}=J_{31}$ and the eigenvalues are
\begin{eqnarray}
E_{\text{D}1}=3U_1-\frac{3J}{4}-J_{23},\\E_{\text{D}2}=3U_1-\frac{3J}{4}-J_{12},
\end{eqnarray}
for which the corresponding wavefunctions are: $|\text{D}_{S_z}\rangle_1$ and $|\text{D}_{S_z}\rangle_2$ [given
by Eq.(\ref{f1}) and (\ref{f2})], respectively. Fig.3 presents the plot of these eigenenergies as a function of the electric field.
In this case
$|\text{D}_{S_z}\rangle_1$ is the dark spin state, but its energy
$E_{\text{D}1}<E_{\text{D}2}$ for $g_E<4t$ only.
Fig.3 also shows the eigenergies $E_{\text{D}1'}$ and $E_{\text{D}2'}$ for the
case $\theta=\pi/3$ (when the electric field is perpendicular to the 13
bond and its direction is opposite to the 2-nd site). Now the corresponding eigenstates are
\begin{eqnarray}
|\text{D}_{S_z}\rangle_{1'}=\frac{1}{\sqrt{2}}\;
  c^{\dagger}_{2\sigma}(c^{\dagger}_{3\sigma}c^{\dagger}_{1\overline{\sigma}}
  -c^{\dagger}_{3\overline{\sigma}}c^{\dagger}_{1\sigma})|\mathrm{vac}\rangle,\\
  |\text{D}_{S_z}\rangle_{2'}=\frac{1}{\sqrt{6}}\;[2c^{\dagger}_{2\overline{\sigma}} c^{\dagger}_{3\sigma}c^{\dagger}_{1\sigma}
  -c^{\dagger}_{2\sigma}(c^{\dagger}_{3\sigma}c^{\dagger}_{1\overline{\sigma}}
  +c^{\dagger}_{3\overline{\sigma}}c^{\dagger}_{1\sigma})]|\mathrm{vac}\rangle.
\end{eqnarray}
However in this case the dark spin state $|\text{D}_{S_z}\rangle_{1'}$ is the
excited state ($E_{\text{D}1'}>E_{\text{D}2'}$). In the both presented cases the
states $|\text{D}_{S_z}\rangle_1$, $|\text{D}_{S_z}\rangle_{1'}$ are
constructed from the singlet states, whereas
$|\text{D}_{S_z}\rangle_2$, $|\text{D}_{S_z}\rangle_{2'}$ are
constructed from the triplet states.

\section{Transport in sequential tunneling regime}

Let us now analyze electronic transport in a sequential tunneling
regime through our system with the left and the right electrode
connected to the QD 1 and 2, respectively. In our
calculations we need transfer rates from the L (R) electrode to the
molecule \cite{Weinmann95}
\begin{eqnarray}\label{tr}
  \Gamma^{L(R)+}_{\nu_2\to{}\nu_3} = \gamma_{L(R)}\, \sum_{\sigma}
  |\langle \nu_3 | c^\dagger_{1(2)\sigma}|\nu_2\rangle|^2 f(\Delta E_{\nu_2\nu_3}-\mu_{L(R)}),\nonumber\\
\end{eqnarray}
where $|\nu_n\rangle$ denotes the initial state for $n$ electrons,
$\Delta E_{\nu_2\nu_3}=E_{\nu_3}-E_{\nu_2}$ is the corresponding
energy difference, $\gamma_{L(R)}$ is a net transfer rate through the
potential barier between the electrode and the molecule, $f$ denotes
the Fermi distribution function, the chemical potentials are taken as
$\mu_L=E_F$, $\mu_R=E_F+eV$, $E_F$ is the Fermi energy and $V$ is a
bias voltage. Similarly, one can write
$\Gamma^{L(R)-}_{\nu_3\to{}\nu_2}$ for the reverse tunneling process
when the electron leaves the molecule.

Since the quadruplet functions $|{\mathrm{Q}}_{S_z}\rangle$ are
constructed from the triplets $|\text{T}_{S_z}\rangle$, therefore, the
nonzero transfer matrix elements are only:
\begin{eqnarray}
|\langle {\mathrm{Q}}_{\pm
  3/2} | c^\dagger_{1(2)\sigma}|{\mathrm{T}}_{\pm
  1}\rangle|=\sqrt{2}\;|\langle {\mathrm{Q}}_{\pm 1/2} |
c^\dagger_{1(2)\sigma}|{\mathrm{T}}_{0}\rangle|\nonumber\\=\sqrt{3}\;|\langle
{\mathrm{Q}}_{\pm 1/2} |
c^\dagger_{1(2)\overline{\sigma}}|{\mathrm{T}}_{\pm
  1}\rangle|=|x^T_{2(1)3}|\,
\end{eqnarray}for $\sigma=\pm 1/2$.
  Here, we use $|{\mathrm{T}}_{S_z}\rangle$ (the singlet solution $|\text{S}\rangle$)
as a linear combination of triplets (singlets) localized on the $ij$
bonds, and $x^T_{ij}$ ($x^S_{ij}$) denote the corresponding
coefficients. The doublet wavefunction is a linear combination of $
|\text{D}_{S_z}\rangle_1$ and $ |\text{D}_{S_z}\rangle_2$ [with the
coefficient $x^D_1$ and $x^D_2$] as well as the states with double
site occupancy. We have checked that for a large $U_0$ the
transfer matrix elements for the states with double occupied sites
play a minor role in electronic transport, and thus they are ignored. The
corresponding elements are:
\begin{eqnarray}
|\langle {\mathrm{D}}_{\pm 1/2} | c^\dagger_{1\overline{\sigma}}|{\mathrm{T}}_{\pm 1}\rangle|=\sqrt{2}\;|\langle {\mathrm{D}}_{\pm 1/2} | c^\dagger_{1\sigma}|{\mathrm{T}}_{0}\rangle|
\nonumber\\ \approx\sqrt{2/3}\;|x^D_2x^T_{23}|,\\
|\langle {\mathrm{D}}_{\pm 1/2} |c^\dagger_{2\overline{\sigma}}|{\mathrm{T}}_{\pm 1}\rangle|=\sqrt{2}\;|\langle {\mathrm{D}}_{\pm 1/2}| c^\dagger_{2\sigma}|{\mathrm{T}}_{0}\rangle|\nonumber\\ \approx|(x^D_1/\sqrt{2}+x^D_2/\sqrt{6})x^T_{13}|,\\
|\langle {\mathrm{D}}_{\pm 1/2}|c^\dagger_{1\sigma}|{\mathrm{S}}\rangle|\approx|x^D_1x^S_{23}|,\\
|\langle {\mathrm{D}}_{\pm 1/2}|c^\dagger_{2\sigma}|{\mathrm{S}}\rangle|\approx 1/2|(x^D_1-\sqrt{3}x^D_2)x^S_{13}|.
\end{eqnarray}

Next, we solve the corresponding master equation~\cite{Weinmann95}
\begin{eqnarray}\label{master}
\frac{dP_{\nu_n}}{dt}&=&
\sum_{\ell,\nu_{n-1}}\Gamma^{\ell+}_{\nu_{n-1}\to{}\nu_n}P_{\nu_{n-1}}+
\sum_{\ell,\nu_{n+1}}\Gamma^{\ell-}_{\nu_{n+1}\to{}\nu_n}P_{\nu_{n+1}}
 \nonumber \\
 &-&P_{\nu_n}
(\sum_{\ell,\nu_{n-1}}\Gamma^{\ell-}_{\nu_n\to{}\nu_{n-1}}+
\sum_{\ell,\nu_{n+1}}\Gamma^{\ell+}_{\nu_n\to{}\nu_{n+1}})
\end{eqnarray}
to find the
occupation probability $P_{\nu_n}$ of the eigenstates $|\nu_n\rangle$
in the steady limit, i.e. for $dP_{\nu_n}/dt=0$.
The current in this limit reads
\begin{equation}\label{current}
I = e\sum_{\nu_2,\nu_3}
\left(\Gamma^{L+}_{\nu_2\to{}\nu_3}P_{\nu_2} -\Gamma^{L-}_{\nu_3\to{}\nu_2}P_{\nu_3}\right).
\end{equation}

\begin{figure}[ht]
\centerline{\epsfxsize=0.47\textwidth \epsfbox{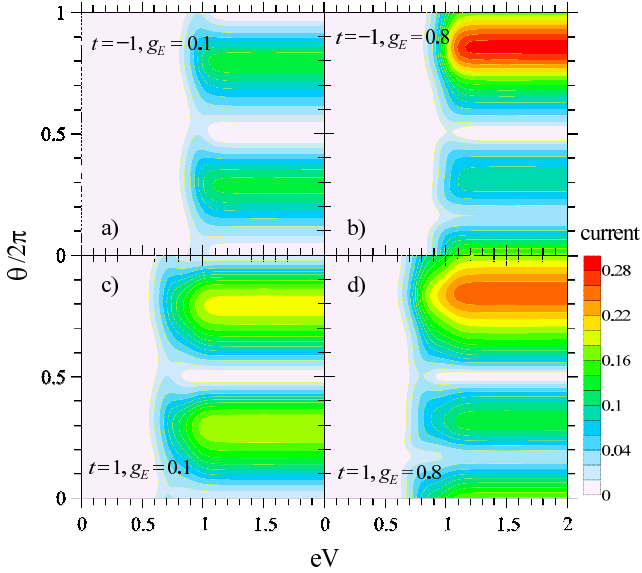} }
\caption{(Color online) Maps of the current as a function of the bias
  voltage $V$ and the angle $\theta$ of the
  electric field with respect to the triangle for $g_E=0.1$ (left column) and
  $g_E=0.8$ (right column). The Fermi level is taken $E_F=5.1$, so at
  equilibrium the system contains two electrons in the singlet
  (triplet) state for $t=-1$ ($t=1$) - top (bottom) plots,
  respectively. We also assumed that $U_0=30$, $U_1=2$, the temperature $T=0.05$, $\gamma_L=\gamma_R=\gamma_0$, and
  the current is in units of $e\gamma_0$.}
\label{map}
\end{figure}

The results of numerical calculations are presented in Fig.\ref{map}
as a map in the $V$-$\theta$ space.  Here we assumed that the
  electric field parameter $g_E$ can be controlled independently of
  the bias voltage, e.g. by means of application of additional
  lateral or back gate electrodes.  For small
fields ($g_E=0.1$) one can see negative differential resistance (NDR)
(an increase and next a drop of $I$ with $V$) at $\theta\approx 0$ and
$\theta\approx \pi$. The NDR effect is due to charge accumulation on
the dark state and the interchannel Coulomb blockade \cite{Emary}. We
have also analyzed all current contributions $I_{\nu_2,\nu_3}$ through
various energy levels. As expected the current flows via the singlet
and the both doublet states for $t=-1$ (top plots in Fig.4). Although the other states are
in the voltage window, they play a minor
role and the corresponding $I_{\nu_2,\nu_3}$ are exponentially small
(these processes are thermally activated only).
The situation for $g_E=0.8$ (Fig.\ref{map}b) is different. One sees that the values of $I$ are higher and
the $\theta$ dependence is different. At $\theta\approx \pi/3$ a new
minimum appears. It is an evidence of the Stark effect, which is also
manifested in activation of the triplet and the
quadruplet states. The transfer of electrons via the quadruplet state is now substantial and its
contribution $I_{\text{T,Q}}$ is larger than those from the other
states. This situation can be explained by analyzing the activation
energies $\Delta E_{\nu_2,\nu_3}$ in Fig.5a and the corresponding transfer rates $\Gamma^{L+}_{\nu_2\to{}\nu_3}$.
Since the singlet is the nondegenerate ground state, it shows the
quadratic Stark effect and $\Delta E_{\text{SD}}$ is a parabola.
In contrast to that the activation energies from the triplet state $\Delta
E_{\text{TD}}$ and $\Delta E_{\text{TQ}}$ show, for
$\theta=5\pi/3$, a linear and a parabolic dependence.
For this case the triplet
levels can be derived explicitly as:
\begin{eqnarray}\label{et}
E_{\text T}=U_1+t-\frac{g_E}{2},\nonumber\\
  E_{\text T}=U_1+\frac{1}{4}(g_E - 2 t \pm \sqrt{9
  g_E^2 + 12 g_E t + 36 t^2}).
\end{eqnarray}
We took the Fermi energy as
$E_F=5.1$, thus, for a small $g_E$, $\Delta E_{\text{TD}}<\Delta
E_{\text{TQ}}<E_F$ and the corresponding transfer rates
$\Gamma^{L+}_{\text{T}\to{}\text{D}}$ and
$\Gamma^{L+}_{\text{T}\to{}\text{Q}}$ [see Eq.(\ref{tr})] are
exponentially small. For larger fields ($g_E>0.2$) these energies are
above $E_F$ and the transfer rates
$\Gamma^{L+}_{\text{T}\to{}\text{D}}$ and
$\Gamma^{L+}_{\text{T}\to{}\text{Q}}$ are activated together with
$\Gamma^{L+}_{\text{S}\to{}\text{D}}$ (the current starts to flow
through all these levels at a threshold voltage).

\begin{figure}[ht]
\centerline{\epsfxsize=0.47\textwidth \epsfbox{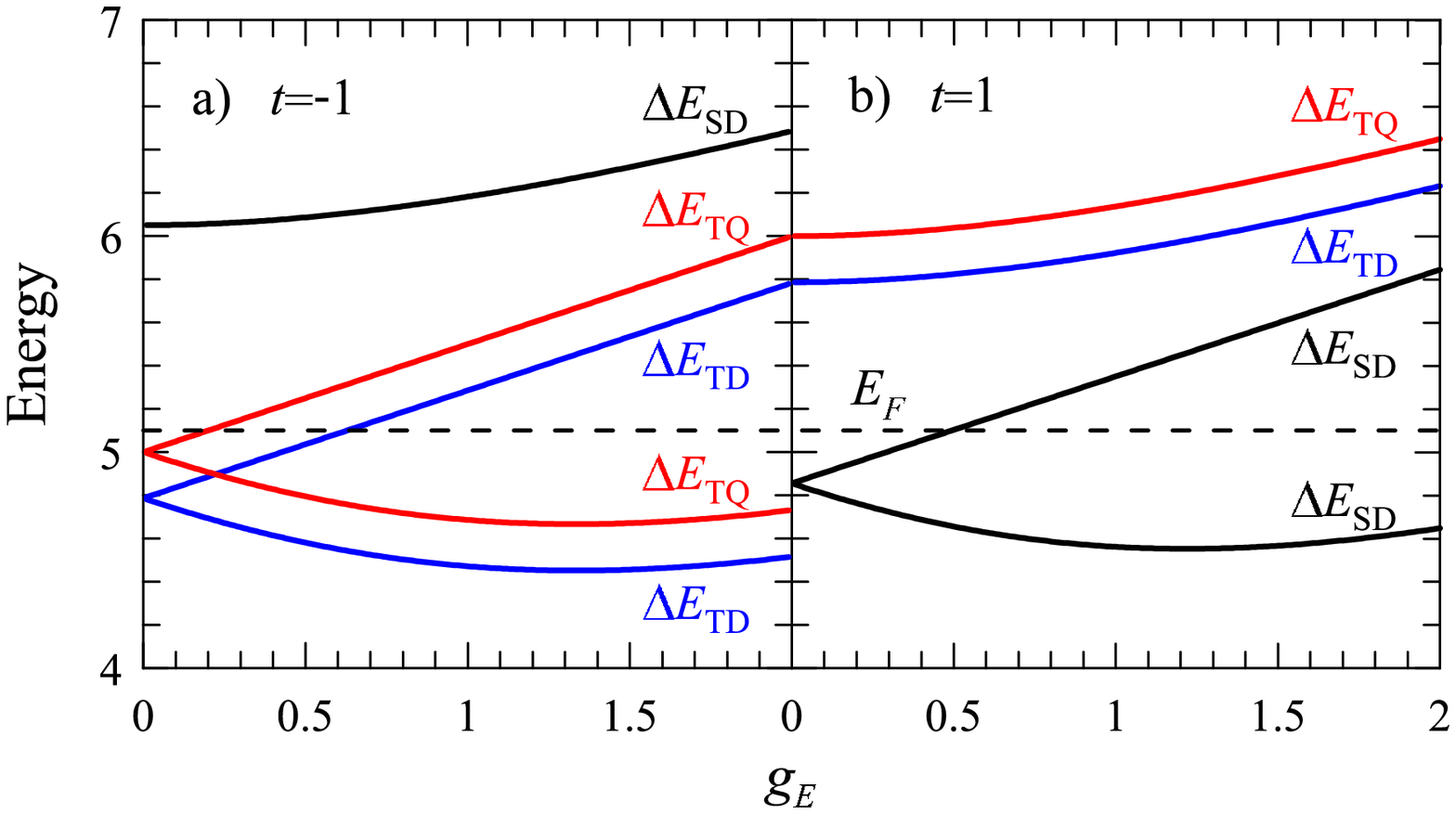} }
\caption{(Color online) Electric field dependence of the energy difference
  $\Delta E_{\nu_2\nu_3}$ for the most relevant states participating
  in electronic transport. The plots a) and b) are for the case $t=-1$ and $t=1$, which correspond to the
  current maps in Fig.4 for the top and the bottom panels at $\theta= 5\pi/3$ and
  $E_F=5.1$, when at equilibrium the ground state is the
  singlet and the triplet, respectively.  Here, we omit the indices 1 and 2 for the doublet states,
  because their activation energies are very close to each other. }
\end{figure}

The bottom row of Fig.4 presents the current maps for the case $t=1$,
for which the triplet is the ground state at equilibrium.
For a bias voltage larger than a threshold one (for $eV\gtrsim 0.7$), electrons are
transferred via the triplet and the doublet states as well as the
quadruplet state. We can clearly see (at $eV\approx 0.9$) the second step in the current,
when the quadruplet state is activated.  In this case, the current maps are also
different for a small and large $g_E$. This results from the dependence of
the activation energies $\Delta E_{\nu_2,\nu_3}$ (and the transfer rates) on the electric field.
Fig.5b shows that $\Delta E_{\text{SD}}$ (almost) linearly increases with $g_E$ and
for large $g_E$ $\Delta E_{\text{SD}}>E_F$. Therefore, the transfer rate $\Gamma^{L+}_{\text{S}\to{}\text{D}}$ is activated.

It is worth noticing that the degenerate states (the triplet $E_T$ for $t=-1$ as well as the singlet $E_S$ for $t=1$) have different dependences on the electric field [see Eq.(\ref{et}), the plots for $\Delta_{\text{TD}}$ and $\Delta_{\text{TQ}}$ in Fig.5a as well as for $\Delta_{\text{SD}}$ in Fig.5b]. One of them is linear vs $g_E$, whereas the second one is nonlinear. This is in contrast to the Stark effect in atomic physics\cite{friedrich}, for which all degenerate states show a linear field dependence in a wide range.

Here we analyzed the electronic transport, in which two- and three-electron
states (with transitions $|\nu_2\rangle \leftrightarrow|\nu_3\rangle$) participated.
By using the electron-hole symmetry of the model (1) one obtains the same results for
transitions $|\nu_3\rangle \leftrightarrow |\nu_4\rangle$ (between the states with three and four
electrons) - provided that one changes the sign of the hopping $t$.

\section{Conclusions}

Summarizing, we considered the influence of the electric field on
strongly correlated electrons in the triangular molecule (the linear and the
quadratic Stark effect).
The orientation $\theta$ of $\mathbf{E}$ with
respect to the molecule is important, because the electric field
breaks the symmetry of the system and changes the symmetry of wave
functions. For some $\theta$ one finds the dark states, responsible
for negative differential resistance.
For $n=3$ electrons we derived the kinetic exchange coupling $J_{ij}$
between the spins, which showed quadratic and linear dependence on
$\mathbf{E}$. The spin-spin correlation functions exhibit different
angle $\theta$ characteristics for a small and large $\mathbf{E}$. In
particular, we studied the dark spin states and their evolution with
$\mathbf{E}$.  The model can be applied to studies of entanglement of
three spin qubits \cite{acin} with the electric field in its specific
role. In particular, it can be applied to a description of an experiment
just recently performed in a system of three QDs,~\cite{laird} which presented
 coherent spin manipulation in a qubit with the logical basis formed from
 the doublet states: $|\text{D}_{S_z}\rangle_1$ and
 $|\text{D}_{S_z}\rangle_2$.
Moreover, we predict that the anisotropic Stark effect
should be seen in electronic transport.

The studied model is general and can be applied to real molecules
with the triangular symmetry, to study their magnetic and optical
features of interest for molecular spintronics. For example, we
predict that spacial anisotropy induced by the electric field in the effective
Heisenberg model will be manifested in molecular magnetism
(e.g. in magnetization, magnetic susceptibility, or ESR spectra)
\cite{gatteschi,choi, tsukerblatt}.  Moreover, the model can be the
paradigm for materials with strongly correlated electrons on triangular
lattices \cite{bulaevskii,cobaltates,powell}. In our opinion
multiferroics are the best candidates to observe the Stark effect,
because in such materials local ferroelectric orderings can modify
exchange couplings between spins as well as magnetic orderings.

\acknowledgments{
We would like to thank Arturo Tagliacozzo for stimulating discussions. This work was supported by Ministry of Science and Higher Education
(Poland) from sources for science in years 2009-2012 and by the EU
project Marie Curie ITN NanoCTM.}

\appendix*

\section{Canonical transformation of the Hubbard model with modulation
  of site energy}

Here we apply a canonical perturbation theory for the Hamiltonian
(\ref{model}) for $U_1=0$, which we rewrite here as:
\begin{equation}\label{H}
H = W + T_0 +T_{-1}+ T_{+1}
\end{equation}
where $W$ represent the sum of all the single site terms, and $T_n$ is
a contribution to the hopping part of the Hamiltonian which increases
by $n$ the number of the double occupied sites in the system. Using
Hubbard operators\cite{Hubbard-64} $X_{i}^{\alpha\alpha'}\equiv|\alpha\rangle\langle\alpha'|$
(defined in terms of the exact eigenstates $|\alpha\rangle$
for an isolated site $i$)
the contributions to the model (\ref{H}) can be
represented by:
\begin{eqnarray}\label{HX}
  W &=& \sum_{i,\alpha}E_{i\alpha}X_{i}^{\alpha\alpha} \nonumber \\
  T_0 &=& \sum_{i<j,\sigma} t_{ij}
  (X_{i}^{\sigma0}X_{j}^{0\sigma}+X_{i}^{2\sigma}X_{j}^{\sigma2}),
  \nonumber\\
  T_{+1} &=& \sum_{i<j,\sigma}\sigma t_{ij}
  X_{i}^{2,-\sigma}X_{j}^{0\sigma},  \hspace{2em}T_{-1}=T^\dagger_{+1}
\end{eqnarray}
The site energies: $E_{i\alpha} \in \{0,\epsilon_{i}, 2\epsilon_{i}+U_0\}$ correspond to the eigenstates
$\alpha\in\{0,\uparrow,\downarrow,2\}$.
The following analysis is valid for a limit
of the small $T_n$, when the differences between energies $E_{i\alpha}$ for
single and double occupied sites are much larger than the hopping parameters.
 In the derivation of the effective Hamiltonian we apply the perturbation
theory with respect to the hopping part, which can be formulated in
terms of the recursive canonical transformation.\cite{macdonald} With
the use of the Hubbard operators the method can be easily generalized
to an arbitrary form of the on\---site zero\---order Hamiltonian,
including the site\---dependent terms.\cite{maska}

\subsection{The effective second order spin Hamiltonian}

Up to the second order with respect to the hopping part the
transformed Hamiltonian reads
\begin{eqnarray}\label{heff}
\tilde{H}=e^{-iS}He^{iS}\approx H_0+\frac{1}{2}[iS,H_1].
\end{eqnarray}
where $H_0=W+T_0$ and $H_1=T_{+1}+T_{-1}$. Here, it is assumed that in
the expansion the linear term with respect to the off\---diagonal part
of hopping vanishes, which is guaranteed by a condition:
\begin{eqnarray}
[iS,H_0]=-H_1.
\end{eqnarray}
From this condition we can derive in the explicit form the
transformation matrix
\begin{eqnarray}\label{S}
iS= \sum_{i<j}iS_{ij}, \nonumber\\ iS_{ij}=
\sum_{\sigma}\left(\frac{ \sigma t_{ij} }{\Delta_{ij}} X_i^{-\sigma2}X_j^{\sigma0}-
\frac{ \sigma t_{ij} }{\Delta_{ji}}
X_i^{\sigma0}X_j^{-\sigma2} \right.\nonumber\\
\left.-\frac{ \sigma t_{ij} }{\Delta_{ji}} X_i^{0\sigma}X_j^{2-\sigma}+
\frac{ \sigma t_{ij} }{\Delta_{ij}} X_i^{2-\sigma}X_j^{0\sigma}\right).
\end{eqnarray}
where $\Delta_{ij}=\epsilon_i+ U_0-\epsilon_j$.
After inserting the operator $S$ (\ref{S}) into (\ref{heff}) we obtain
the effective Hamiltonian valid up to second order perturbation with
respect to the hopping part.
In a form projected to the subspace
$C_{00}$, defined as subspace of many-electron states with all sites singly
occupied, i.e. for $n=3$ electrons
in the triangle, the Hamiltonian reads:
\begin{eqnarray}\label{Heff-X}
  \tilde{H}|_{C_{00}}&=& W+
  \frac{1}{2}\sum_{i<j,\sigma,\alpha}\sigma\alpha\,J^{(2)}_{ij}
  \,X_i^{-\sigma\alpha}X_j^{\sigma,-\alpha}
\end{eqnarray}
where
\begin{eqnarray}
J^{(2)}_{ij} = 2t_{ij}^2(\Delta_{ij}^{-1}+\Delta_{ji}^{-1}).\end{eqnarray}
The effective Hamiltonian can be rewritten in a more familiar form
with a help of the spin operators
\begin{eqnarray}
\tilde{H}|_{C_{00}} &=& W+ \sum_{i<j}J^{(2)}_{ij}
\left(\vec{S}_i\cdot\vec{S}_j- \frac{1}{4}
\right).
\end{eqnarray}

\subsection{The effective third order spin Hamiltonian}
A derivation of the higher order terms in a general case is based on
the recursive procedure,\cite{macdonald} however it is rather involved
and will be discussed in a separate paper. Here we only present the
extra 3rd order term projected to the $C_{00}$ subspace
\begin{eqnarray}
\left.\tilde{H}^{(3)}\right|_{C_{00}} &=&
\sum_{i<j} J^{(3)}_{ij}\left(\vec{S}_i\cdot\vec{S}_{j}-\frac{1}{4}\right).
\end{eqnarray}
The exchange parameter reads:
\begin{eqnarray}
\label{J3}
J^{(3)}_{ij} =
2\,t_{ji}t_{im}t_{mj}
\left(\Delta^{-1}_{ij}\Delta^{-1}_{im}+\Delta^{-1}_{ji}\Delta^{-1}_{jm}\right.\nonumber\\ + \Delta^{-1}_{im}\Delta^{-1}_{jm} \left.-\Delta^{-1}_{ji}\Delta^{-1}_{mi}-\Delta^{-1}_{ij}\Delta^{-1}_{mj}
  -\Delta^{-1}_{mi}\Delta^{-1}_{mj}
\right).
\end{eqnarray}
Here, the indices $i,j,m$ denote three different sites. Note, that the extra term vanishes for the uniform case $\epsilon_1=
\epsilon_2= \epsilon_3$.

By expanding $J^{(2)}_{ij}$ and $J^{(3)}_{ij}$ [Eqs.(A.7) and (A.10)] in a series with respect to $\epsilon_i\ll U_0$ we ontain the exchange coupling Eq.(5).

\end{document}